\documentclass[showpacs,preprintnumbers]{revtex4}
\usepackage{graphicx}
\usepackage{amssymb}

\begin{document}
\title{Kerr black holes in horizon-generating form}
\author{Sean A. Hayward}
\affiliation{Department of Science Education, Ewha Womans University,
Seodaemun-gu, Seoul 120-750, Korea\\ {\tt hayward@mm.ewha.ac.kr}}
\date{28th January 2004}

\begin{abstract}
New coordinates are given which describe non-degenerate Kerr black holes in
dual-null foliations based on the outer (or inner) horizons, generalizing the
Kruskal form for Schwarzschild black holes. The construction involves an area
radius for the transverse surfaces and a generalization of the Regge-Wheeler
radial function, both functions of the original radial coordinate only.
\end{abstract}
 \maketitle

It is hard to overstate the importance of the Kerr solution either in the
theoretical history of black holes or as a model for actual astrophysical black
holes. The solution was discovered \cite{Ker} before black holes were
understood as such, but it soon became a key ingredient in the classical theory
of black holes \cite{BCH,HE,MTW,Wal}. Uniqueness theorems indicate that this
small family of solutions, parametrized by just mass and angular momentum,
constitute all stationary black holes in vacuum, and therefore serve as good
models for most real black holes for most of the time, from supernova remnants
to supermassive black holes mysteriously found in galactic cores. Dynamical
black holes have a partial theory being currently developed
\cite{bhd,1st,in,cyl,ql1,gwbh,mg9,AK1,AK2,BF}, for which an appropriate
understanding of Kerr black holes will also be necessary.

The generic Kerr metric is much more complex than the Schwarzschild metric, to
which it reduces in the non-rotating case, and indeed Kerr discovered it almost
a half-century after Einstein formulated General Relativity \cite{Ein}, as
compared to Schwarzschild's few weeks \cite{Sch}. While it has deservedly been
one of the most studied exact solutions, there appears to be a peculiar lack of
natural coordinates generating the outer (or inner) horizons, as in the Kruskal
extension \cite{Kru} of the Schwarzschild solution, which was so crucial for
understanding its nature as a black hole. (There is an earlier paper of
Fronsdal \cite{Fro} which acknowledges priority to unpublished work of Kruskal;
the Kruskal paper was actually written by Wheeler \cite{Whe}). Specifically,
recall that a non-degenerate Kerr black hole has two asymptotically flat
external regions and two outer Killing horizons, intersecting in a bifurcation
surface or wormhole throat \cite{BL,Car}. Kruskal's coordinates (or rather
their sum and difference) describe a dual-null foliation of the space-time,
meaning two families of null hypersurfaces intersecting in a two-parameter
family of transverse spatial surfaces, such that the horizons are two of the
hypersurfaces. Such coordinates are naturally adapted to the horizons and
radiation propagation, both into the black hole and out of its vicinity. This
Letter reports generalizations to non-degenerate Kerr black holes.

The Kerr solution with mass $m$ and angular momentum $ma$ in Boyer-Lindquist
\cite{BL} coordinates $(t,r,\theta,\phi)$ can be given by the inverse metric
\begin{equation}
g^{-1}=\frac1\Sigma\left(\Delta\partial_r^2 +\partial_\theta^2
+\frac{\Delta-a^2\sin^2\theta}{\Delta\sin^2\theta}
\partial_\phi^2 -\frac{4mar}\Delta
\partial_\phi\partial_t-\frac\Pi\Delta\partial_t^2\right)
\end{equation}
where $\partial_x$ are the coordinate derivative vectors,
$\partial_x(dx')=\delta_{xx'}$, symmetric tensor products are implied and
\begin{eqnarray}
\Delta&=&r^2+a^2-2mr\\
\Sigma&=&r^2+a^2\cos^2\theta\\
\Pi&=&(r^2+a^2)^2-\Delta a^2\sin^2\theta.
\end{eqnarray}
The inverse metric is crucial in this adventure, since one is seeking null
coordinates $x^\pm$: $g^{-1}(dx^\pm,dx^\pm)=0$. For $a<m$, the Killing horizons
are located at $r=r_\pm$, where $r_\pm=m\pm\sqrt{m^2-a^2}$ are the roots of
$\Delta(r)=0$. Here the outer horizons $r=r_+$ will be studied in detail, with
modifications for the inner horizons $r=r_-$ mentioned below. The desired
transformation may be given in two steps, firstly to coordinates
$(t^*,r^*,\vartheta,\varphi)$ defined by
\begin{eqnarray}
t^*&=&t-a\sin\theta\label{newt}\\
r^*&=&\int\frac{R^2}\Delta dr\label{newr}\\
\vartheta&=&\theta\\
\varphi&=&\phi-\Omega(t-a\sin\theta)\label{newphi}
\end{eqnarray}
where $\Omega=a/2mr_+$ is the angular velocity of the horizon and
\begin{equation}
R(r)=\left((r^2+a^2)^2-a^2\Delta\right)^{1/4}=\left(r^4+a^2r^2+2ma^2r\right)^{1/4}.
\end{equation}
Secondly one may transform to $(x^+,x^-,\vartheta,\varphi)$, where
\begin{equation}
x^\pm=\pm e^{\kappa(r^*\pm t^*)}
\end{equation}
and $\kappa$ is determined below to be the surface gravity. The last step
constructs dual-null coordinates $x^\pm$ by a generalization of Kruskal's
method, having first identified appropriate coordinates $(t^*,r^*)$. The
colatitude $\vartheta$ has been renamed merely for clarity in partial
derivatives, and the new longitude $\varphi$ can be partly understood by noting
that it is normal to the horizon-generating Killing vector
$\chi=\partial_t+\Omega\partial_\phi$: $\chi(d\varphi)=\chi(d\vartheta)=0$. The
same is true for $\phi-\Omega t$ alone \cite{Cha}, but it ceases to be spatial
at large radius. The key step is the identification of the new coordinate
$t^*$, which absorbs mischievous $\theta$-dependent terms to leave the
corresponding $r^*$ (such that $r^*\pm t^*$ are null) miraculously available as
a function of $r$ alone. One may recognize $r^*$ as a generalization of the
Regge-Wheeler \cite{RW} tortoise coordinate, named after the ancient paradox of
Achilles and the tortoise, as $r=r_+$ corresponds to $r^*\to-\infty$.

The coordinate derivative vectors are related by
\begin{eqnarray}
\partial_t&=&
\kappa(x^+\partial_+-x^-\partial_-)-\Omega\partial_\varphi\\
\partial_r&=&\frac{R^2}\Delta\kappa
(x^+\partial_++x^-\partial_-)\\
\partial_\theta&=&\partial_\vartheta+\Omega a\cos\vartheta\partial_\varphi
-\kappa a\cos\vartheta(x^+\partial_+-x^-\partial_-)\\
\partial_\phi&=&\partial_\varphi
\end{eqnarray}
where $\partial_\pm=\partial_{x^\pm}$. Here $r$ is retained as a function of
$x^+x^-=-e^{2\kappa r^*}$, determined implicitly by inverting (\ref{newr}) to
give $r(r^*)$. By carefully arranged cancellations in $R^4+\Delta
a^2\cos^2\theta-\Pi=0$, the inverse metric transforms to
\begin{equation}
g^{-1}=\frac1\Sigma\left(\frac{4R^4\kappa^2x^+x^-}{\Delta}\partial_+\partial_-
+2\kappa(x^+\partial_+-x^-\partial_-)
(\alpha\partial_\varphi-a\cos\vartheta\partial_\vartheta) +\partial_\vartheta^2
+2\Omega a\cos\vartheta\partial_\vartheta\partial_\varphi
+\left(\frac1{\sin^2\vartheta}-\beta\right)\partial_\varphi^2\right)
\end{equation}
where
\begin{eqnarray}
\alpha(r)&=&\frac{\Omega R^4-2mar}\Delta\\
\label{beta}\beta(r)&=&\frac{\Omega^2 R^4-4\Omega mar+a^2}\Delta.
\end{eqnarray}
Note that $x^\pm$ are null by the absence of $\partial_\pm^2$ components, and
that $r=r_+$ has been mapped to $x^+x^-=0$. Then $x^\pm=0$ are the two branches
of the outer horizons, with $x^+=x^-=0$ being the bifurcation surface. The
Killing vector $\chi=\kappa(x^+\partial_+-x^-\partial_-)$ has a saddle point at
each angle of the bifurcation surface: it vanishes there, and generates the two
branches of the horizons along the principal directions. Having constructed a
dual-null foliation based on the horizons in particular coordinates, the
remaining coordinate freedom consists of relabelling the null hypersurfaces,
$x^\pm\mapsto\hat x^\pm(x^\pm)$, and diffeomorphisms of a transverse surface,
i.e. choice of angular coordinates.

The functions $(\alpha,\beta,x^+x^-/\Delta)$, each apparently singular at
$\Delta=0$, are all actually regular at $r=r_+$. Firstly, since
$\Delta=(r-r_+)(r-r_-)$, near $r=r_+$ one has
$r^*\sim(R^2/(r-r_-))|_{r=r_+}\ln(r-r_+)$, so $e^{2\kappa r^*}\sim r-r_+$ if
\begin{equation}
\kappa=\left.\frac{r-r_-}{2R^2}\right|_{r=r_+}=\frac{\sqrt{m^2-a^2}}{2mr_+}
\end{equation}
which is the usual surface gravity. Then $x^+x^-/\Delta\sim-1/(r-r_-)$ is
regular at $r=r_+$. Secondly one can rearrange
\begin{eqnarray}
\alpha&=&\Omega\left(\frac{(r+r_+)(r^2+a^2)}{r-r_-}+r_+^2\right)\\
\beta&=&\Omega^2\left(\frac{(r+r_+)^2(r-r_+)}{r-r_-}+2r_+^2+a^2\right)
\end{eqnarray}
which are now manifestly regular at $r=r_+$. Then $g^{-1}$ is regular, except
for the polar coordinate breakdown at $\sin\vartheta=0$. With this usual
caveat, the coordinate chart covers the entire region outside the inner
horizons.

For the inner horizons, one can use the same formulae
(\ref{newt})--(\ref{beta}) if one replaces the angular velocity and surface
gravity by $\Omega=a/2mr_-$ and $\kappa=-\sqrt{m^2-a^2}/2mr_-$ respectively.
Thus it is convenient to use a negative surface gravity for the inner horizons,
as is anyway natural for dynamical horizons \cite{1st,in,cyl,ql1,gwbh,mg9}.
Then the new coordinate chart covers the region inside the outer horizons, up
to the discs $r=0$, where $R=0$. Apart from the ring singularities at
$\Sigma=0$, the discs are regular and one can extend through to the internal
asymptotically flat regions, which can also be covered by a similar chart. The
ring singularities present topological obstructions to any encircling foliation
by spheres.

The metric
\begin{equation}
g=\frac\Sigma\Delta dr^2+\Sigma d\theta^2+\frac1\Sigma\left(\Pi\sin^2\theta
d\phi^2-4mar\sin^2\theta d\phi dt-(\Delta-a^2\sin^2\theta)dt^2\right)
\end{equation}
in the new coordinates may be found by inverting the coordinate transformation
to give the coordinate differentials
\begin{eqnarray}
dt&=&dt^*+a\cos\vartheta d\vartheta\\
dr&=&\frac\Delta{R^2}dr^*\\
d\theta&=&d\vartheta\\
d\phi&=&d\varphi+\Omega dt^*.
\end{eqnarray}
An independent and more systematic method uses the general dual-null
decomposition \cite{dne}
\begin{equation}
g^{-1}=-2e^f\partial_+\partial_- +2e^f(s_+\partial_-+s_-\partial_+)-2e^fs_+s_-
+h^{-1}
\end{equation}
where $h^{-1}$ is the inverse metric of the (constant-$x^\pm$) transverse
surfaces, $s_\pm$ are shift vectors and $f$ is a normalization function. In
this case one finds
\begin{eqnarray}
e^f&=&-\frac{2R^4\kappa^2x^+x^-}{\Delta\Sigma}\\
s_\pm&=&\pm\frac{\Delta}{2R^4\kappa x^\pm}
(\alpha\partial_\varphi-a\cos\vartheta\partial_\vartheta)\\
h^{-1}&=&\frac1{R^4\Sigma}\left(\Pi\partial_\vartheta^2
+4ma^2r\cos\vartheta\partial_\vartheta\partial_\varphi
+\left(\frac{R^4}{\sin^2\vartheta}-a^2r(r+2m)\right)\partial_\varphi^2\right).
\end{eqnarray}
The determinant of $h^{-1}$ is just $1/R^4\sin^2\vartheta$, so it may be easily
inverted to yield the transverse metric
\begin{equation}
h=\frac1\Sigma\left((r^4+r(r+2m)a^2\cos^2\vartheta)d\vartheta^2
-4ma^2r\cos\vartheta\sin^2\vartheta d\vartheta d\varphi+\Pi\sin^2\vartheta
d\varphi^2\right)
\end{equation}
where symmetric tensor products are again implied. Then the metric is given by
the general form \cite{dne}
\begin{equation}
g=h+2h(s^+)dx^++2h(s^-)dx^-+h(s_+,s_+)(dx^+)^2+h(s_-,s_-)(dx^-)^2
+2(h(s_+,s_-)-e^{-f})dx^+dx^-
\end{equation}
by calculating
\begin{eqnarray}
h(s_\pm)&=&\pm\frac\Delta{2\Sigma\kappa x^\pm}\left((\alpha+\Omega
a^2\cos^2\vartheta)\sin^2\vartheta d\varphi
-(1+\gamma\sin^2\vartheta)a\cos\vartheta d\vartheta\right)\\
h(s_\pm,s_\pm)&=&\frac\Delta{\Sigma(2\kappa x^\pm)^2}
\left((\beta+\Omega^2a^2\cos^2\vartheta)\sin^2\vartheta-1+\Sigma^2/R^4\right) \\
h(s_+,s_-)&=&-\frac\Delta{4\Sigma\kappa^2x^+x^-}
\left((\beta+\Omega^2a^2\cos^2\vartheta)\sin^2\vartheta-1+\Sigma^2/R^4\right)
\end{eqnarray}
where
\begin{equation}
\gamma(r)=\Omega\alpha-\beta=\frac{2\Omega mar-a^2}\Delta =\frac{2\Omega
ma}{r-r_-}.
\end{equation}
The terms involving the shift vectors are quite complex as compared to those in
the inverse metric. However, the transverse metric $h$ has an unexpectedly nice
property: its area form is just $R^2\sin\vartheta d\vartheta\wedge d\varphi$.
(All the usual coordinates \cite{Ker,BL,Car,Cha} have a transverse area form
with more complex $\theta$-dependence). Thus the new radial function $R$ is
revealed as the area radius of the transverse surfaces. This property is useful
for evaluating surface integrals, most simply the area $4\pi R^2$ of the
transverse surfaces. Also, while the coordinates were intended only to cover
the outer horizons, the angular coordinates $(\vartheta,\varphi)$ become
standard spherical polar coordinates on round spheres as $R\to\infty$:
$h/R^2\to d\vartheta^2+\sin^2\vartheta d\varphi^2$.

One might at first think that this dual-null foliation is unique, since the
foliations of each horizon are generated by $\chi$ from the bifurcation
surface. However, $\chi$ vanishes at the bifurcation surface, leaving the
possibility to include it in different foliations of the horizons. In fact,
another such foliation can be found simply by replacing $a$ with $-a$ in the
definitions of $t^*$ (\ref{newt}) and $\varphi$ (\ref{newphi}), which has the
effect of changing the sign of $d\vartheta$ in the metric, or of
$\partial_\vartheta$ in the inverse metric. The two sets of transverse surfaces
have the same poles, but their equators are displaced relative to each other.
The displacements are in opposite directions in the two external regions, so
the two foliations are dual under left-right inversion of the space-time. The
existence of such dual foliations is due to breaking the explicit symmetry
$(t,\phi)\mapsto(-t,-\phi)$ in $(t^*,\varphi)$. One may choose whether to
displace the time coordinate to the future or past in one universe, forcing the
opposite choice to be made in the other universe. Despite this peculiarity, it
seems likely that this pair of foliations is unique in some sense. For
instance, one might conjecture that they are the only dual-null foliations
based on the horizons for which the transverse area form is a function of $r$
times the area form of a unit sphere.

The new coordinates are so naturally adapted to Kerr horizons and radiation
propagation that they are likely to have various applications, while the area
radius $R$ and the generalized tortoise coordinate $r^*$ may themselves prove
useful. Applications might include electromagnetic radiation powered by
rotating black holes, gravitational radiation produced by perturbed black
holes, quasi-normal modes, Cauchy-horizon instability and black-hole
evaporation. It might also help in the search for practical definitions of
energy-momentum and angular momentum in the absence of symmetries, instead
using extrinsic properties of embedded surfaces. The author's main interest in
this enterprise was to find the Kerr solution in a similar dual-null form to
that which has proved useful in understanding dynamical black holes. Its role
as a basic test case may now be employed to further develop a general theory of
black-hole dynamics.

\medskip\noindent Supported by research grant ``Black holes and gravitational
waves'' of Ewha Womans University.

\end{document}